\documentstyle[preprint,eqsecnum,epsfig,prd,aps,floats]{revtex}
\begin{document}


\newcommand{\stln}{\setlength{\unitlength}{2.0ex}}
\newcommand{\smln}{\setlength{\unitlength}{1.0ex}}
\newcommand{\fr}{\framebox(1,1){}}
\newcommand{\sfr}{\framebox(1,1){\begin{picture}(1,1)
  \put(0,0){\line(1,1){1}}\end{picture}}}

\newcommand{\onebox}
{\lower0.3\unitlength \hbox{
\begin{picture}(1.6,1.6)
\put(.1,.3){\fr}
\end{picture}}}

\newcommand{\twobox}
{\lower0.3\unitlength\hbox{
\begin{picture}(2.6,1.6)
\put(.1,.3){\fr} \put(1.1,.3){\fr}
\end{picture}}}

\newcommand{\oneonebox}
{\lower0.9\unitlength\hbox{
\begin{picture}(1.6,2.6)
\put(.1,.3){\fr} \put(.1,1.3){\fr}
\end{picture}}}

\newcommand{\twoonebox}
{\lower0.8\unitlength\hbox{
\begin{picture}(2.6,2.6)
\put(.1,1.3){\fr} \put(1.1,1.3){\fr} \put(0.1,0.3){\fr}
\end{picture}}}

\newcommand{\twotwobox}
{\lower0.8\unitlength\hbox{
\begin{picture}(2.6,2.6)
\put(.1,.3){\fr} \put(.1,1.3){\fr} \put(1.1,.3){\fr}
\put(1.1,1.3){\fr}
\end{picture}}}

\newcommand{\oneoneonebox}
{\lower1.4\unitlength\hbox{
\begin{picture}(1.6,3.6)
\multiput(.1,.3)(0,1){3}{\fr}
\end{picture}}}

\newcommand{\twooneonebox}
{\lower1.4\unitlength\hbox{
\begin{picture}(2.6,3.6)
\multiput(.1,.3)(0,1){3}{\fr} \put(1.1,2.3){\fr}
\end{picture}}}

\newcommand{\oneoneoneonebox}
{\lower2.1\unitlength\hbox{
\begin{picture}(1.6,4.6)
\multiput(.1,.3)(0,1){4}{\fr}
\end{picture}}}

\newcommand{\twooneoneonebox}
{\lower2.1\unitlength\hbox{
\begin{picture}(2.6,4.6)
\multiput(.1,.3)(0,1){4}{\fr} \put(1.1,3.3){\fr}
\end{picture}}}

\newcommand{\twotwooneonebox}
{\lower2.1\unitlength\hbox{
\begin{picture}(2.6,4.6)
\multiput(.1,.3)(0,1){4}{\fr} \multiput(1.1,2.3)(0,1){2}{\fr}
\end{picture}}}

\newcommand{\twotwoonebox}
{\lower1.4\unitlength\hbox{
\begin{picture}(2.6,3.6)
\multiput(.1,.3)(0,1){3}{\fr} \put(1.1,1.3){\fr}
\put(1.1,2.3){\fr}
\end{picture}}}

\newcommand{\twotwotwobox}
{\lower1.4\unitlength\hbox{
\begin{picture}(2.6,3.6)
\multiput(.1,.3)(0,1){3}{\fr} \multiput(1.1,.3)(0,1){3}{\fr}
\end{picture}}}

\newcommand{\sonebox}
{\lower0.3\unitlength\hbox{
\begin{picture}(1.6,1.6)
\put(.1,.3){\sfr}
\end{picture}}}

\newcommand{\stwobox}
{\lower0.3\unitlength\hbox{
\begin{picture}(2.6,1.6)
\put(.1,.3){\sfr} \put(1.1,.3){\sfr}
\end{picture}}}

\newcommand{\soneonebox}
{\lower0.9\unitlength\hbox{
\begin{picture}(1.6,2.6)
\put(.1,.3){\sfr} \put(.1,1.3){\sfr}
\end{picture}}}


\newcommand{\nl}{\nonumber \\}
\newcommand{\one}{$(1,0)$}
\newcommand{\zz}{$Z_2$}
\newcommand{\ee}{$E_8$}
\newcommand{\osp}{$OSp\,(6,2|\,2)$}
\newcommand{\isom}{$SU(2)_R \times SU(2)_L$}
\newcommand{\ba}{\begin{array}}
\newcommand{\ea}{\end{array}}

\preprint{\vbox{\hbox{CALT-68-2208}
                \hbox{hep-th/XXXXXXX}  }}

\title{\Large The Operator Spectrum of the Six-dimensional $(1,0)$ Theory }

\author{Eric G. Gimon\thanks{email: egimon@theory.caltech.edu} and
    Costin Popescu\thanks{email: popescu@theory.caltech.edu}}

\address{California Institute of Technology, Pasadena, CA 91125}

\maketitle

\begin{abstract}
We study the large N operator spectrum of the (1,0) superconformal
chiral six-dimensional theory with $E_8$ global symmetry.  This
spectrum is dual to the Kaluza-Klein spectrum of supergravity on
$AdS_7 \times S^4/Z_2$ with a ten-dimensional $E_8$ theory at its
singular locus.  We identify those operators in short multiplets
of $OSp\,(6,2|\,2)$, whose dimensions are exact for any N.  We
also discuss more general issues concerning AdS/CFT duality on
orbifold supergravity backgrounds.
\end{abstract}

\newpage
{
\tighten

\section{Introduction}

    In the last year important new methods for studying superconformal
theories have been developed based on a conjecture by
Maldacena~\cite{malda}. This conjecture relates D-dimensional
superconformal theories to string theory/M-theory to
$AdS_{D+1}\times M$, where $M$ is a compact manifold. Using a
prescription given by Gubser, Klebanov and Polyakov~\cite{BKP} and
Witten~\cite{witten1} one can associate operators on the
D-dimensional boundary with fields in the bulk of $AdS_{(D+1)}$
and compute correlation functions for these operators using
supergravity.

    In the spirit of much of the modern work in string duality, our
goal in this paper is not the presentation of further {\it
evidence} for the AdS/CFT correspondence, but rather to {\it use}
this correspondence to determine the unknown spectrum of operators
for interacting superconformal theories.  Of particular interest
for us is the six-dimensional (1,0) fixed point theory with $E_8$
global symmetry~\cite{ganorhanany,seibergwitten}.  One feature
which makes this theory particularly interesting is that it has
both a Coulomb branch along which states charged under $E_8$
decouple, and a Higgs branch with broken $E_8$ global symmetry.
The transition from one branch to the other can be used to connect
four-dimensional theories with different numbers of chiral
fields~\cite{kachrusilverstein}.   Unfortunately this otherwise
useful field theory has no known Lagrangian description
(see~\cite{seiberg}).  There does exist a discrete light-cone
quantized description for the \one\
theory~\cite{abks,lowe,johnson}, but this does not include the
operator spectrum.  In order to obtain better control over the
properties of the \one\ superconformal fixed point we use AdS/CFT
duality to extract its spectrum of chiral operators.

     To derive the spectrum of chiral operators, we
determine the spectrum of Kaluza-Klein modes for the M-theory
background in the large N limit where supergravity is
valid~\footnote{We label by N the \one\ theory which can be
resolved to N/2 free tensor theories in its Coulomb branch}. The
K--K modes couple to gauge invariant operators on the AdS boundary
with dimensions which are either protected or receive 1/N
corrections, and provide a description of the chiral operator
spectrum of the $(1,0)$ theory which remains useful for small N.
Other six-dimensional $(1,0)$ theories (with different global
symmetries) have been studied in~\cite{ferrara,tatar}).

    Finding the supergravity background appropriate for describing
the \one\ theory involves orbifolding the dual background for the
interacting large N $(2,0)$ six-dimensional theory (studied in the
AdS/CFT context by~\cite{Oz-p,Minwalla-p,leigh,Halyo-p}) . The
orbifold process brings interesting new aspects to the AdS/CFT
story (see for example~\cite{shamiteva,juanetal}). In our analysis
we bring out new features such as the appearance of multiple types
of short superconformal multiplets, generic to all orbifolds which
break the CFT R-symmetry.

    The paper is organized as follows.  In section II, we describe
the singular M-theory geometry conjectured in~\cite{berkooz} which
yields the AdS dual to the \one\ theory.  Fields in this geometry
as well as the operators of the \one\ theory fall into
representations of $OSp\,(6,2|\,2)$.  In Section III we build
short multiplets of interest for this supergroup using the
oscillator construction of~\cite{oscillator}.  Section IV matches
$E_8$ neutral operators of the \one\ theory with K--K reduced
fields from the bulk M-theory geometry, while section V matches
operators charged under $E_8$ with the K--K reduction of fields
living at the singular locus of this geometry. Finally in section
5 we summarize by listing relevant and marginal operators of the
\one\ theory, and comment on their physical relevance.  We also
discuss at some length some general features of orbifold
constructions.

\section{Compactification Geometry}

    The \one\ theory we are interested in can be thought of as placing a
large number, $N$, of M5-branes on top of an ``end-of-the-world''
9-branes of the Ho\v{r}ava-Witten compactification of
M-theory~\cite{Horava},
 $R^{1,9}\times S^1/Z_2$.  The near horizon geometry of N M5-branes is
 $AdS_7 \times S^4$, where the $S^4$ is described by the equation
\begin{equation}
\label{sphere} (X^6)^2 + (X^7)^2 + (X^8)^2 + (X^9)^2 + (X^{11})^2
= R^2.
\end{equation}
R is a fixed radius whose size grows with N.

    Since the $Z_2$ mentioned above flips the sign of $X^{11}$,
it is natural to believe that large N \one\ theory is dual to
M-theory on $AdS_7 \times S^4/Z_2$ as conjectured
in~\cite{berkooz}. The $Z_2$ action has a fixed point locus,
$AdS_7 \times S^3$, on which propagates the $D=10$ ${\cal N}=1$
SYM $E_8$ multiplet derived in the Ho\v{r}ava-Witten
compactification. This $E_8$ twisted sector was present in the
M-theory background before we took the near-horizon limit. It can
also be directly motivated, as the gravitational anomaly
argument~\cite{Horava} which dictates it's presence at $Z_2$
singularities in flat space still applies in our curved
background.

    Upon reduction of M-theory down to the $AdS_7$,
there are two kinds of K--K modes.  The K--K modes of the first
kind are given in terms of the K--K modes of M-theory on $S^4$
which survive the $Z_2$ projection. These modes carry no $E_8$
quantum numbers and hence couple only to $E_8$-neutral operators
on the boundary of $AdS_7$. The K--K-modes from the fixed $S^3$
have adjoint $E_8$ quantum numbers and couple to charged operators
on the $AdS_7$ boundary, in a manner similar to the one described
in~\cite{juanetal}. Note that since only adjoint $E_8$ fields
exist in the bulk, any operator on the boundary in a
representation of $E_8$ other than the adjoint will have to couple
to a multi-particle state~\footnote{See~\cite{deboer} for more on
the role of multi-particle states.}.

    Two more points need to be made about the modes on this M-theory
background.  First, the isometry group, $SO(5)$, of the original
$S^4$ is broken to $SO(4)=SU(2)_R \times SU(2)_L$ by the $Z_2$
projection.  The gauge group for $AdS_7$ is then $SU(2)_R \times
SU(2)_L \times E_8$, but we pick a basis such that only the first
factor is embedded in the superconformal algebra \osp . Second,
the ``twisted sector'' \ee\ fields will not shift the large N mass
spectrum of the K--K bulk modes since they only affect tree-level
computations when they are given a non-zero vev. For large N the
size of the $S^4$ grows much faster than that of the fixed point
$S^3$, therefore masses for bulk Kaluza-Klein modes unprotected by
supersymmetry will receive at most 1/N corrections from
interactions with the twisted sector fields.

\section{Oscillator Construction of \osp\ Short Multiplets}
\label{construct}

    Before going on to calculate the mass spectrum of the
low-energy Kaluza-Klein M-theory modes on the geometry described
above, we would like to first take a moment to construct the short
multiplets of \osp\ which will be relevant to our problem. We do
this because any K--K modes which we can fit into one of these
short multiplets will have mass eigenvalues fixed by the \osp\
group theory, making explicit calculations unnecessary. We follow
the methods of G\"unaydin etal.~\cite{oscillator} used for
constructing multiplets of $OSp(6,2|4)$ with just a slight
modification to allow for the reduced R-symmetry.  Since this
method is extensively described in~\cite{oscillator}, we will only
give a brief review.

    The basic idea is to start with the maximal compact proper subgroup of
the bosonic part of the \osp\ superalgebra, $SO(6,2) \times
SU(2)_R$. This subgroup is $U(4)_B \times U(1)_F$.  We next
introduce $p$ pairs of bosonic creation/annihilation operators
$a^i(r) = (a_i(r))^{\dag}$ and $b^i(r) = (b_i(r))^{\dag}$
transforming in the $4$ and ${\bar 4}$ representation of $U(4)_B$
(with $r = 1,\ldots,p$), as well as a pair of fermionic
creation/annihilation operators $\alpha ^{\dag}(r),\alpha (r)$ and
$\beta ^{\dag}(r), \beta (r)$ with $U(1)_F$ charges $+{1 \over 2}$
and $-{1 \over 2}$ respectively.  We organize these oscillators
into column vectors:
\begin{equation}
\label{columns}
\begin{array}{lll}
\xi_A(r) =
\left(
\begin{array}{c} a_i(r) \\ \alpha(r) \end{array}
\right) \,\;
& \eta_A(r) =
\left(
\begin{array}{c} b_i(r) \\
\beta(r)
\end{array}
\right), &  \\ &&\\ \xi^A(r) = \left(\begin{array}{c} a^i(r) \\
\alpha^{\dag}(r) \end{array}\right),\; & \eta^A(r) =
\left(\begin{array}{c} b^i(r) \\ \beta^{\dag}(r)
\end{array}\right),\; &
\left\{
\begin{array}{l}
r = 1,\ldots,p, \\ i,j,\ldots=1,\ldots,4
\end{array}
\right.
\end{array}.
\end{equation}
($A = \cdot$ will represent the one fermionic index) whose only
non-zero commutation/anti-commutation relations can be represented
symbolically as
\begin{equation}
\label{relations}
 \Big\{ \xi_A(r),\xi ^B(s) \Big] = \delta_A^{\
\,B} \delta^r_s \qquad \Big\{ \eta_A(r),\eta ^B(s) \Big] =
\delta_A^{\ \,B} \delta^r_s.
\end{equation}

        The Lie superalgebra can now be realized in terms of
bilinears in $\xi$ and $\eta$.  They are given by
\begin{equation}
\label{generators}
\begin{array}{l}
A_{AB} = \xi_A \cdot \eta_B - \eta_A \cdot \xi_B, \\ \\ A^{AB} =
A^{\dag}_{AB} = \eta^B \cdot \xi^A - \xi^B \cdot \eta^A, \\ \\
M^A_{\ \,B} = \xi^A \cdot \xi_B + (-1)^{deg(A)deg(B)}\eta_B \cdot
\eta^A ,
\end{array}
\end{equation}
where $deg(A)=0$ for a bosonic index $A$ and $deg(A)=1$ for a
fermionic index $A$.  The even subalgebra $SO(6,2) \times SU(2)_R$
is generated by the elements \{$A_{ij}, A^{ij}, M^i_{\ j}$\} and
\{$ A_{\displaystyle{\cdot\cdot}}, A^{\displaystyle{\cdot\cdot}},
M^{\displaystyle \cdot}_{\ \,\displaystyle \cdot}$\}.

    This oscillator construction of generators naturally
implements a Jordan decomposition of \osp\ with respect to the
maximal compact subgroup $U(4|1)$ graded with the $U(1)$
generator, $Q = {1 \over 2} M^{A}_{\ \,A}$:
\begin{equation} \label{decomposition}
L = L^{-}\oplus L^0 \oplus L^+.
\end{equation}
The generators $A_{AB}$ and $A^{AB}$ correspond to the $L^-$ and
$L^+$ spaces, respectively.  The generators $M^A_{\ \,B}$ of
$U(4|1)$ give the $L^0$ space.  We use this Jordan decomposition
to generate unitary irreducible representations (UIR's) of \osp\ .
We start with a lowest weight state, $|\,\Omega \rangle$, in an
irreducible representation of $L^0$ and annihilated by all the
generators in $L^-$. The complete UIR is then generated using
successive applications of generators in $L^+$ on $|\,\Omega
\rangle$.

    To understand the physical interpretation of states in a
given UIR, it is convenient to further decompose representations
of $U(4|1)$ into representations of $U(4)_B \times U(1)_F$. We
label the $U(4|1)$ representations in terms of super-Young
tableaux~\footnote{These super-Young tableaux are presented in
detail within section 5 of \cite{oscillator}.}. Their
decompositions look like: \stln
\begin{equation}\label{stableau}
{\soneonebox} \to (\!\oneonebox,1),\; (\!\onebox,\!\onebox),\;
(1,\!\twobox).
\end{equation}

The group $U(4)_B \simeq Spin(6)\times U(1)_B$ is the maximal
compact subgroup of $SO(6,2)$.  $U(1)_B$ is generated by the
charge $Q_B = {1 \over 2} M^i_{\;i} = {1 \over 2} (N_B + 4p)$.  In
the $AdS$/CFT duality, this charge corresponds to the $AdS$ energy
of a supergravity mode and to the dimension of its dual CFT
operator. $N_B$ is the bosonic number operator, and for a $U(4)$
representation is just the number of boxes in the corresponding
Young tableau.  For that given Young tableau, the $Spin(6)$
representation can be recovered by matching the $SU(4)$ indices
with the appropriate $SU(4)$-invariant tensor.  A few
examples\footnote{More details can be found at the end of section
2 of~\cite{oscillator}} are:
\begin{equation}\label{names}
{{\textstyle |\;0\;\rangle\;,} \atop \!\mbox{scalar}} \qquad
{{\textstyle |\onebox\;\rangle\;,} \atop \!\mbox{spinor}}\qquad
{{\textstyle |\twobox\;\rangle\;,} \atop \!
{\mbox{A}}_{\alpha\beta\gamma}} \qquad {{\textstyle
|\oneonebox\;\rangle\;,} \atop \! \mbox{vector}}
\end{equation}
Finally, $U(1)_F$ is generated by $Q_F = {1 \over 2}
M^{\displaystyle \cdot}_{\;\displaystyle \cdot} = {1 \over 2}(N_F
- p)$, which measures $SU(2)_R$ spin.

    Now that we have a better understanding of how to interpret
states in representations of $U(4|1)$, let us quickly describe the
action of the $L^+$ generators.  In order to do this, we break up
the set $L^+$ into the subsets:~\begin{equation}
L^+ =
\left\{A^{AB}\right\} \to \left\{
\begin{array}{lll}
   B^+ = \{A^{ij}\}
\\ Q^+ = \{A^{i\displaystyle \cdot}\}
\\ F^+ = \{A^{\displaystyle{\cdot\cdot}}\}
\end{array}
\right. .
\end{equation}
Given a representation of $U(4)_B\times U(1)_F$, $B^+$ gives
conformal descendants.  In AdS these correspond to higher energy
Fourier-like modes.  $F^+$ generates the complete set of spin
states in a given $SU(2)_R$ representation. Finally, $Q^+$ is the
set of supersymmetry generators.

    To illustrate how the oscillator construction works, we start
with the simplest example
\begin{equation}\label{simple}
|\,\Omega \rangle = |\, 0\rangle_p,
\end{equation}
the vacuum corresponding to $p$ pairs of oscillators.  This state
is a scalar with dimension/energy equal to $2p$ and lowest
$SU(2)_R$ spin $-{p \over 2}$.  With its $F^+$ descendants,
$|\,\Omega\rangle$ fills out the {\bf(p+1)} of $SU(2)_R$. Acting
with $Q$ gives the following supermultiplet of conformal primary
states~(lowest energy states): \smln
\renewcommand{\arraystretch}{1.3}
\begin{equation}\label{chiral}
\begin{array}{|c|r|c|c|} \hline
U(4)          & \mbox{State}\           & SU(2)_R & \Delta
\\ \hline\hline  1  & \mbox{Scalar} & {\bf(p+1)}    & \,2p
\\ \hline\onebox    & \mbox{Spinor}        & {\bf p}       & \,(2p + {1 \over 2})
\\ \hline\oneonebox & \mbox{Vector}        & {\bf (p-1)}   & \,(2p + 1)
\\ \hline\oneoneonebox &\mbox{Spinor}        & {\bf (p-2)}   & \,(2p + {3 \over 2})
\\ \hline\oneoneoneonebox &\mbox{Scalar}        & {\bf (p-3)}   & \,(2p +2) \\ \hline
\end{array}
\end{equation}
\renewcommand{\arraystretch}{1}
For $p < 4$ we keep only those states with positive $SU(2)_R$
dimension.  For readability's sake we always include the $Spin(6)$
representation labeled in terms of fields in $AdS_7$.

    The highest spacetime spin in the multiplet (\ref{chiral}) is the
vector, indicating a short representation of the superalgebra. In
fact, if we act with $Q$ on the superconformal primary, the
topmost scalar, its highest weight $SU(2)_R$ component is
annihilated.
This means that we are dealing with a level one short multiplet in
the formalism discussed by~\cite{minwalla}. The superconformal
primary state in this type of short multiplet (the topmost scalar)
has dimension $\Delta = 4s$ where $s$ is the $SU(2)_R$ spin.  We
can check this explicitly:
\begin{equation} \label{dimension}
\Delta = Q_B |\,0\rangle_p = 2p = -4\, Q_F |\,0\rangle_p = 4s.
\end{equation}
In order to differentiate from other short multiplets, we call
this type of multiplet a chiral multiplet. It will play an
important role in our analysis.

    There is another short multiplet of \osp\ that we will find
useful, starting with the ground state
\begin{equation}
\label{newground} \xi^{[A} \xi^{B]} \, |\,0\rangle_p.
\end{equation}
This ``ground state", with super-Young tableau given in
(Eq.~\ref{stableau}), gives a vector of dimension $(2p+1)$, a
spinor of dimension $(2p + {1 \over 2})$ and a scalar of dimension
$2p$. Since the scalar has lowest dimension, we will refer to it
as the superconformal primary.  The states in this multiplet are:
\stln
\renewcommand{\arraystretch}{1.3}
\begin{equation}\label{other}
\begin{array}{|c|c|c|c|}
\hline U(4) & \mbox{State}\,   & SU(2)_R        & \Delta
\\ \hline \hline
0           &   \mbox{Scalar}\,      & {\bf (p-1)}   & \, 2p
\\ \hline
\onebox     &   \mbox{Spinor}\,      & \ba{c} {\bf  p   } \\ {\bf
                (p-2)} \ea           & \,(2p + {1 \over 2})
\\ \hline
\twobox     &   \mbox{3-Form}\,      & {\bf (p-1)}   & \,(2p + 1)
\\ \hline
\oneonebox  &   \mbox{Vector}\,      & \ba{c} {\bf (p+1)} \\
                {\bf (p-1)}          \\ {\bf (p-3)} \ea & \,(2p + 1)
\\ \hline
\twoonebox  & \,\mbox{Gravitino}\,   & \ba{c} {\bf  p} \\ {\bf
                (p-2)} \ea           & \,(2p + {3 \over 2})
\\ \hline
\twotwobox  &   \mbox{Graviton}\,    & {\bf (p-1)}   & \,(2p + 2)
\\ \hline
\oneoneonebox&  \mbox{Spinor}\,      & \ba{c} {\bf  p   }\\
                {\bf (p-2)} \\ {\bf (p-4)}\ea & \,(2p + {3 \over 2})
\\ \hline
\end{array}
\qquad
\begin{array}{|c|c|c|c|}
\hline U(4) &   \mbox{State}\,       & SU(2)_R        & \Delta
\\ \hline \hline
\twooneonebox&  \mbox{2-Form}\,      & \ba{c} {\bf (p-1)} \\
                {\bf (p-3)} \ea       & \,(2p + 2)
\\ \hline
\twotwoonebox &\,\mbox{Gravitino}\,  & \ba{c} {\bf (p-2)} \\
                {\bf p} \ea          & (2p + {5 \over 2})
\\ \hline
\oneoneoneonebox & \mbox{Scalar}\,   & \ba{c} {\bf (p-1)} \\
                {\bf (p-3)} \\ {\bf (p-5)} \ea & \,(2p + 2)
\\ \hline
\twooneoneonebox & \mbox{Spinor}\,   & \ba{c} {\bf (p-2)} \\
                    {\bf (p-4)} \ea  & \,(2p + {5 \over 2})
\\ \hline
\twotwooneonebox &\mbox{Vector}\,    & {\bf (p-3)}
                 & \,(2p+ 3)
\\ \hline \multicolumn{4}{l}{} \\
\end{array}
\end{equation}
\renewcommand{\arraystretch}{1}
Again, we only allow states with positive $SU(2)_R$ dimension.
Even then, for $p<5$ some of the elements listed above are not in
fact conformal primaries, but $F^+$ and $B^+$ descendants of other
primaries.  For example, when $p=2$ the vector in the {\bf 1} of
$SU(2)_R$ is the $F^+$ descendant of the vector in the {\bf 3}.
This odd behavior is related to the fact that this multiplet does
not develop null states until level 3.  The superconformal primary
(scalar) for this short multiplet has $\Delta = 4s + 4$.  We know
from~\cite{minwalla} that short multiplets of \osp\ can have
scalar superconformal primaries with
\begin{equation}\label{listshort}
\Delta = 4s,\;4s+2,\;4s+4,\;4s+6,
\end{equation}
but only the ones described in this chapter will be relevant for
the supergravity spectrum.

\section{Kaluza-Klein Reduction of the Bulk Eleven-Dimensional Supergravity}

    Now that we have established some of the multiplet structure
of \osp\ we can carry out our analysis of the bulk $S^4$
Kaluza-Klein modes which survive the $Z_2$ projection.  The group
theory for this process is simple.  These K--K modes come from
$S^4$ harmonics in representations of $SO(5)$.  Decomposing $SO(5)
\to SU(2)_R\times SU(2)_L \times Z_2$ , with $Z_2= {\scriptstyle
+/-}$ for harmonics which are even/odd under $X^{11} \to -X^{11}$,
we reduce all the 11-dimensional supergravity fields on the even
harmonics except for the 3-form~\footnote{Here we refer to forms
in index free notation}. This last field is reduced on odd
harmonics since it flips sign under parity reversal. For
convenience, we will describe the K--K spectrum in terms of the
dual CFT operators and apply the $Z_2$ projection on the CFT
spectrum.

    The $S^4$ K--K spectrum derived in Ref.~\cite{massless,KKspectrum}
can be nicely organized in terms of dual operators using
``place-holder" fields~\cite{FF}. In~\cite{leigh} the short
multiplet operator spectrum for the $(2,0)$ theory dual to
$AdS_7\times S^4$ was written in terms of place-holder scalar,
spinor and tensor fields
\begin{equation}\label{placeholders}
\tilde\phi,\: \tilde\psi,\: \tilde H
\end{equation}
taken to be in the adjoint of $U(N)$.  They transform in the {\bf
5}, {\bf 4}, and {\bf 1} representation of $SO(5)$~\footnote{We
denote with a \~\  fields in $SO(5)$ R-symmetry representations}.
Starting with a superconformal primary operator
\begin{equation}
{\cal O}_{\,0,\,0,\,p} = Tr\: \tilde\phi^p
\end{equation}
(the power $p$ is schematic, the $\tilde\phi$'s are actually in
symmetric traceless representation of $SO(5)$) we build a complete
supermultiplet of conformal primary operators
\begin{equation}
{\cal O}_{\,m,\,n,\,k} = Tr\: \tilde
H^m\tilde\psi^n\tilde\phi^k,\quad m+n+k=p.
\end{equation}
Note that in the oscillator method used to generate superconformal
multiplets of $OSp\,(6,\!2|\,4)$, these multiplets are built using
the vacuum ground state with the same number $p$ of oscillator
flavors (see~\cite{oscillator}). The p=1 operators Tr~$\tilde H$,
Tr~$\tilde\psi$, and Tr~$\tilde\phi$ are the only operators in the
Abelian part of $U(N)$ and correspond to the doubleton.  They
decouple from the theory. The $p=2$ supermultiplet (also referred
to as the massless multiplet) is very important, as it contains
the R-symmetry currents, the super-currents and the stress-energy
tensor, as well as relevant scalar operators.

        To look at the operators dual to the K--K modes of $S^4/Z_2$
we use a simple extension of the methods above.  We split the
place-holder fields into an even group and an odd group.  The even
group contains scalars and spinors
\begin{equation}\label{even}
\ba{ll}
    \phi\quad &\mbox{in the}\;{\bf (2,2)},
\\  \psi\quad &\mbox{in the}\;{\bf (1,2)}, \ea
\end{equation}
where $SU(2)_R \times SU(2)_L$ is now the global symmetry. These
place-holders transform in the anti-symmetric ({\bf AS}) of
$USp(N)$ (N must be even) and loosely correspond to fluctuations
inside the end-of-the-world 9-brane. Since we have broken $U(N)$
to $USp(N)$, traces of place-holder fields include the symplectic
matrix~{\bf J}.  This means any number of these even place-holder
fields can appear in a trace ($Tr[\,{\bf J} \cdot \mbox{\bf AS}]
\neq 0$).

The odd group contains a scalar, spinors, and a self-dual 2-form
\begin{equation}\label{odd}
\ba{ll}
    \rho\quad & \mbox{in the} \: {\bf(1,1)},
\\  \chi\quad & \mbox{in the} \: {\bf(2,1)}, \: \mbox{and}
\\  H \quad   & \mbox{in the} \: {\bf(1,1)}.
\ea
\end{equation}
These place-holders transform in the adjoint of $USp(N)$, so only
operators represented by traces with an {\em even} number of these
survive the $Z_2$ projection due to the commutation relation of
adjoint $USp(N)$ matrices with {\bf J}.
    For a given $p$, the superconformal primaries ${\cal
O}_{\,0,\,0,\,p}$ of the ${\cal N}=2$ algebra break up into
separate superconformal primaries of the ${\cal N}=1$ algebra. The
ones which survive the $Z_2$ projection can be schematically
written as
\begin{equation}
Tr\:\phi^p,\;Tr\:\phi^{p-2}\rho^2,\;Tr\:\phi^{p-4}\rho^4,\;\ldots
\end{equation}
The first of these primaries will transform in the {\bf(p+1,p+1)}
of \isom\, the next in the {\bf(p$-$1,p$-$1)}, and so on.  They
all inherit the dimension of ${\cal O}_{\,0,\,0,\,p}$, so will
have $\Delta = 2p$. Relating their R-symmetry spin, $s$, to this
dimension we get the relations:
\begin{equation}
\label{listprimary}
\Delta= 4s,\;\Delta=4s+4,\;\Delta=4s+8,\ldots
\end{equation}
For each $p$, the first operator in this series is a
superconformal primary for a chiral short multiplet with content
as in (\ref{chiral}).  The second is also an ${\cal N}=1$
superconformal primary for a short multiplet, but now with the
primary fields listed in (\ref{other}).  The rest of the operators
in (\ref{listprimary}) are superconformal primaries for {\em long}
multiplets of \osp. This can be seen both from their dimension,
scalars with $\Delta > 4s + 6$ are superconformal primaries for
long multiplets only, and from the number of states in the
multiplet. Their dimensions are valid only in the large N limit,
since they can receive corrections of order 1/N when tree-level
supergravity computations are no longer protected by
supersymmetry.

    For $p=1$, only one superconformal operator can be written
down, Tr~$\phi$.  The conformal primaries in this multiplet are
this scalar in the {\bf(2,2)} of $SU(2)_R\times SU(2)_L$ and a
spinor in the {\bf(1,2)}.  The anti-symmetric tensor of $USp(N)$
is reducible, its symplectic trace is a singlet and won't mix with
other operators. This matches with our understanding of the $p=1$
operators as dual to the ``pure gauge" (i.e., no physical degrees
of freedom) AdS fields in the doubleton representation of \osp.
They correspond to the center of mass motion of the M5-branes
parallel to the 9-branes and decouple~\footnote{This decoupling of
the center of mass degrees of freedom follows from coordinate
invariance inside the 9-brane.}.

    We would like to point out that a clear distinction should be
drawn between place-holder fields and doubleton degrees of
freedom.  In the ${\cal N}=2$ formalism, the doubleton and the
place-holder multiplet carry the same quantum numbers, so an
identification might be drawn.  For ${\cal N}=1$ only the even
multiplet of place-holders yields a doubleton, the odd
place-holders are necessary for computing the correct spectrum but
can't appear alone in a trace.  Thus, the pure-gauge doubleton
degrees of freedom of the AdS background should really only be
interpreted as decoupled center of mass degrees of freedom for the
CFT and bear no direct connection to the place-holder fields.

    The odd place-holders first play a role for the $p=2$
operators.  The ${\cal N}=2$ super-primary operator
Tr~$\tilde\phi\tilde\phi$ splits into Tr~$\phi\phi$,
Tr~$\phi\rho$, and Tr~$\rho\rho$.  Only the first and third of
these survive the $Z_2$ projection.  We list the results in the
following table:
\renewcommand{\arraystretch}{1.3}
\begin{equation}
\label{maintable}
\begin{array}{|l|c|ccccc|ccccc|}
\hline \mbox{Type} & \Delta & \multicolumn{5}{c|}{SO(5) \to
SU(2)_R\times SU(2)_L\times Z_2}&
\multicolumn{5}{c|}{\mbox{Place-Holders}}
\\ \hline\hline
\mbox{Scalar} & 4 & {\bf 14} \to &{\bf (3,3)_+} &+&{\bf
(1,1)_+}&+\,\mbox{$Z_2$-odd} & Tr\,\tilde\phi\tilde\phi\to
&Tr\,\phi\phi &+& Tr\,\rho\rho &+\,\mbox{$Z_2$-odd}
\\ \hline\mbox{Spinor} & 4{1\over 2} & {\bf 16} \to &{\bf (2,3)_+}
&+&{\bf (2,1)_+}&+\,\mbox{$Z_2$-odd} & Tr\,\tilde\phi\tilde\psi\to
&Tr\,\phi\psi &+& Tr\,\rho\chi &+\,\mbox{$Z_2$-odd}
\\ \hline\mbox{Vector} & 5 & {\bf 10} \to &{\bf (1,3)_+} &+&{\bf
(3,1)_+}&+\,\mbox{$Z_2$-odd} & Tr\,\tilde\psi\tilde\psi\to
&Tr\,\psi\psi &+& Tr\,\chi\chi &+\,\mbox{$Z_2$-odd}
\\ \hline\mbox{3-Form} & 5 & {\bf 5} \to & & &{\bf (1,1)_+} &+\,\mbox{$Z_2$-odd}
& Tr\,\tilde\phi\tilde H\to & & & Tr\,\rho H &+\,\mbox{$Z_2$-odd}
\\ \hline\mbox{Gravitino} & 5{1\over 2} & {\bf 4} \to & & & {\bf (2,1)_+}
&+\,\mbox{$Z_2$-odd} & Tr\,\tilde\psi\tilde H\to & & & Tr\,\chi H
&+\,\mbox{$Z_2$-odd}
\\ \hline\mbox{Graviton} & 6 & {\bf 1} \to & &
&{\bf (1,1)_+}&+\,\mbox{$Z_2$-odd} & Tr\,\tilde H\tilde H\to & & &
Tr\,H H &+\,\mbox{$Z_2$-odd}\\ \hline
\end{array}
\end{equation}
\renewcommand{\arraystretch}{1}
From this table, we can see that we get two massless multiplets
for $AdS_7\times S^4/Z_2$.  The first is a chiral multiplet and
contains the $SU(2)_L$ gauge field.  The second multiplet contains
the $SU(2)_R$ gauge field, the gravitino, and the graviton.  These
fields couple to the R-symmetry current, the super-current and the
stress-energy tensor of the boundary CFT respectively.  The
presence of the R-symmetry current and the stress-energy tensor in
the same multiplet fixes the relation between R-spin,$\, s$, and
the dimension,\ $\Delta$, of the short multiplets superconformal
primaries~\footnote{The fact that the various short multiplets
relate $\Delta$ and $s$ differently is due to the appearance of
null-states at different levels.}.

    The rest of the $(1,0)$ spectrum dual to the $AdS_7\times
S^4/Z_2$ bulk fields can be easily obtained using the place-holder
fields described above.  The group theory for projecting down
${\cal N}=2$ multiplets is straightforward and this part of our
analysis requires no new understanding of the K--K reduction of
11-dimensional supergravity.  Because the complete Kaluza-Klein
spectrum on $S^4$ has already been worked out
in~\cite{massless,KKspectrum}, we only need to identify those bulk
states which survive the $Z_2$ projection. It's important to note
that since we do not have a Lagrangian UV description for the
\one\ theory, the place-holder formalism is only a labeling system
convenient for doing group theory.

\section{Kaluza-Klein Reduction of $E_8$ Twisted Sector Modes}

    To get the mass spectrum of the Kaluza-Klein reduction of the
$E_8$ twisted sector modes, one might expect to have to calculate
the eigenvalues of the Laplace operator for the $E_8$ ${\cal N}=1$
$D=10$ vector multiplet reduced on a sphere.  Luckily, group
theory comes to our rescue again, in a fashion quite similar to
that in the analysis of~\cite{juanetal}. Upon reduction to $AdS_7$
the $D=10$ vector multiplet can give only scalars, spinors, and
vectors. This implies that its K--K modes must fit into chiral
short multiplets, since all other multiplets include a larger set
of bosonic fields.

     Looking at just the bosonic modes, we derive the complete
Kaluza-Klein spectrum. Our only initial bosonic field is the
$D=10$ vector field. It can be reduced on the fixed $S^3$ using a
scalar harmonic in the {\bf (k,k)} of $SU(2)_R\times SU(2)_L$ or
using a vector harmonic in the {\bf (k+2,k)} or {\bf (k,k+2)}.
These harmonics give a bosonic spectrum of the form
\begin{equation}\label{listchirals}
\ba{lcllllll} \mbox{Scalars}&\to& {\bf (3,1)},\: & {\bf(4,2)},\: &
{\bf (5,3)},\: &  {\bf (6,4)},\: & \ldots
\\ \mbox{Vectors} &\to& {\bf (1,1)},\: &{\bf (2,2)},\: & {\bf (3,3)},\:
& {\bf (4,4)},\: & \ldots
\\ \mbox{Scalars} &\to& & & {\bf (1,3)},\: & {\bf (2,4)},\: & \ldots
\ea
\end{equation}
The columns in this series fit easily into chiral multiplets with
superconformal primary scalars in the {\bf (k+2,k)}.  The $AdS$
energy/dimensions of all these states can be directly read off
from (\ref{chiral}). Just as in~\cite{juanetal}, scalars coming
from the vector harmonics {\bf (k+2,k)} and {\bf (k,k+2)} have
different $AdS$ energies due to a subtlety involving the
eigenvalues of the $E_8$ covariant derivative on $S^3$.

The first multiplet in the series contains the massless vector
which carries the $E_8$ gauge symmetry and couples to the dual
$E_8$ current algebra on the $AdS_7$ boundary CFT. Since this
vector comes from a straight reduction on $S^3$, its coupling is
proportional to
\begin{equation}\label{coupling}
1/\sqrt{Vol(S^3)} \simeq 1/\sqrt{N}.
\end{equation}

\section{Discussion}

    We conclude by giving a summary of chiral gauge invariant
relevant and marginal operators for the \one\ theory.  We make
some general remarks on orbifolds in the AdS/CFT context, and
offer some perspective on the Coulomb/Higgs branch transition.

    We denote operators by their $SU(2)_R \times SU(2)_L \times
E_8$ quantum numbers.  The relevant and marginal scalar operators
are:
\begin{equation}
\begin{array}{ll}
{\bf (3,3,1), (1,1,1), (3,1,248)}\qquad & \mbox{with}\; \Delta =
4\;(relevant) \\ {\bf (4,4,1), (2,2,1), (4,2,248)}\qquad &
\mbox{with}\; \Delta = 6\;(marginal).
\end{array}
\end{equation}
All these scalar operators break supersymmetry since they are
superconformal primary operators and thus are not annihilated by
any supersymmetry transformations.  Also, from the analysis
of~\cite{distlerzamora} for the relevant scalars of the $(2,0)$
theory, we don't expect the {\bf (3,3,1)} and {\bf (1,1,1)}
operators to lead to any new {\it stable} superconformal fixed
points. On the other hand, the {\bf (3,1,248)} scalar operator
will very likely be implicated in breaking the $(1,0)$ theory to
its Higgs branch, as any $E_8$ instanton bundle on the fixed $S^3$
will excite the corresponding Kaluza-Klein mode.  This analysis
requires a solution of the coupled twisted sector and bulk modes
of $AdS_7 \times S^4/Z_2$ and is beyond the scope of our
discussion.

    The rest of the relevant and marginal spectrum behaves as
expected.  The vector spectrum yields all the expected dimension
five global symmetry currents for $SU(2)_R \times SU(2)_L \times
E_8$. We also find the super-current and the stress-energy tensor
and a relevant 3-form current.  An important check on our analysis
is that only the $SU(2)_R$ current appears in the same multiplet
as the stress-energy tensor.  Since this is the only R-symmetry
current, anything else would violate the supersymmetry algebra.
Continuing with this logic, the $SU(2)_L$ and $E_8$ symmetry
currents need to appear in a different kind of short multiplet
(one without the stress-energy tensor). Clearly, differentiation
in the types of short multiplets must appear in any situation
involving CFT's with global symmetries that are not R-symmetries.

    One way to understand the appearance of new short multiplets
for orbifolds of AdS/CFT configurations is by looking more closely
at the role of R-symmetry in the shortening process.  For example,
in the $(2,0)$ spectrum we start with a superconformal primary
operator in the symmetric-traceless of the $SO(5)$ R-symmetry.
Only the highest weight state in this representation is
annihilated by the supersymmetry transformation, but this is
enough to shorten the superconformal multiplet.  When we orbifold
to get the \one\ theory this single $SO(5)$ representation breaks
into multiple surviving $SU(2)_R \times SU(2)_L$ representations.
If we rank these representations in order of decreasing $SU(2)_L$
dimension, the first contains the highest weight state of $SO(5)$
and shortens at level 1.  The next (surviving) representation has
a state which annihilates at level 3, and so is the primary for a
new kind of short multiplet.  Since we can't apply the
supersymmetry generators an infinite number of times (they are
fermionic), a large enough initial $SO(5)$ representation for the
superconformal primary will spawn superconformal primaries for
multiplet that never shorten, i.e., long multiplets.

    The appearance of these long multiplets in the Kaluza-Klein reduced
supergravity single particle spectrum of orbifolds is in itself an
interesting occurrence.  At tree level the states in the long
multiplets inherit their masses from their $(2,0)$ theory
progenitors. For large N, these will then be the correct masses,
but not necessarily for small N.  By contrast, in the standard
story for AdS/CFT duality, with maximally symmetric spaces, single
particle supergravity states are mapped strictly to operators in
short multiplets, i.e., with fixed dimensions.  In the large N
limit where supergravity is valid only these operators keep finite
dimensions, all other operators have anomalous dimensions that
grow with N.  In orbifolds, the new long multiplets provide us
with an extra set of single particle operators, whose naive
dimension becomes increasingly {\it correct} for large N. This is
the kind of behavior usually associated with multi-trace operators
dual to multiple particle supergravity states, where the
interaction energy correction to the naive mass estimate vanishes
in the large N limit (the coupling goes to zero).

    The appearance of exotic multiplets in the single particle
spectrum of orbifolds can also be related to the introduction of
separate sets of place-holder fields for the dual CFT.  In section
IV, we separated place-holders into an even and an odd set.  The
shortest multiplet (denoted chiral) had all even place-holders,
the next shortest had two odd place-holders, and so on.  We expect
this to generalize to other orbifolds as follows.  The
place-holder fields of the covering space theory will be grouped
by their transformation properties under the defining orbifold
quotient operation. Preserved operators will correspond to
invariant combinations of these place-holder fields.  If more than
one type of invariant combination exists (e.g., even fields, or
odd-odd fields), different types of multiplets should ensue.  Note
that only place-holder fields which are singlets under the
orbifold operation yield doubleton fields in AdS.  Therefore, the
naive connection between place-holders and doubletons seen in
maximally symmetric spaces breaks down.

    Returning to specifics of the \one\ theory, we would like to
take a look at the flat directions discussed in the introduction.
One problem with the notion of moduli space inherent in that
discussion is that it is usually parameterized in terms of
expectation values of fields. In the conformal field theory
language we can only perturb our vacuum by adding operators to the
theory.  The coefficients for these additional operators then act
as parameters.  Motion along one of the flat directions of moduli
space typically corresponds to adding marginal operators which are
not gauge invariant, and as such were not analyzed in our
discussion.

    Despite this problem it is still possible to look at the
moduli space of the \one\ theory in the near vicinity of the
interacting superconformal fixed point. The idea is to move a
short distance along moduli space before taking the near-horizon
limit.   For example, if we separate the stack of N M5-branes into
two stacks of N/2 M5-branes, we can get a multi-center AdS space
describing motion along the Coulomb branch.  Due to the nature of
scaling limit, motion away from the 9-branes will always be
limited to be less than $l_P$, so recovery of the $(2,0)$ theory
of N/2 M5-branes is an infinite distance away in moduli space.
Similarly, one can hope to analyze the Higgs branch by
``blowing-up" $E_8$ instantons on the fixed point $S^3$.  This
involves solving coupled twisted sector/bulk equations on $AdS_7
\times S^4/Z_2$, and presents an interesting problem in itself.
Once again, the scaling limit which yields the AdS/CFT
correspondence only allows us to travel a short distance along
this flat direction.

\acknowledgments We would like to thank P.~Ho\v{r}ava,
C.V.~Johnson, S.~Kachru, A.~Lawrence, J.~Maldacena, J.~Minahan,
D.~Minic, S.~Minwalla, and J.~Schwarz for discussions and helpful
comments. This work was supported in part by the U.S.\ Dept.\ of
Energy under Grant no.\ DE-FG03-92-ER~40701. }


\begin{thebibliography}{99}

\bibitem{malda}
J. Maldacena, {\sl The large $N$ limit of superconformal field
theories and supergravity}, Adv.Theor.Math.Phys.{\bf
2}~(1998)~231, hep-th/9711200.

\bibitem{BKP}
S. S. Gubser, I. R. Klebanov and A. M. Polyakov, {\sl Gauge theory
correlators from non-critical string theory}, Phys.Lett.{\bf B428}
(1998) 105, hep-th/9802109.

\bibitem{witten1}
E. Witten, {\sl Anti de Sitter space and holography},
Adv.Theor.Math.Phys.{\bf 2} (1998) 253, hep-th/9802150.

\bibitem{ganorhanany}
O. Ganor and A. Hanany, {\sl Small $E_8$ Instantons and
Tensionless Noncritical Strings}, Nucl. Phys. {\bf B474} (1996)
122, hep-th/9602120.

\bibitem{seibergwitten}
N. Seiberg and E. Witten, {\sl Comments on String Dynamics in Six
Dimensions}, Nucl. Phys. {\bf B471} (1996) 121, hep-th/9603003.

\bibitem{kachrusilverstein}
S. Kachru and E. Siverstein, {\sl Chirality-Changing Phase
Transitions in 4D String Vacua}, Nucl. Phys. {\bf B504} (1997)
272, hep-th/9704185.

\bibitem{seiberg}
N.~Seiberg, {\sl Notes on Theories with 16 Supercharges},
Nucl.Phys.Proc.Suppl.{\bf 67} (1998) 158-171, hep-th/9705117;
N.~Seiberg, {\sl Nontrivial Fixed Points of the Renormalization
Group In Six-Dimensions}, Phys.Lett.{\bf B390} (1997) 169,
hep-th/9609161.

\bibitem{abks}
O. Aharony, M. Berkooz, S. Kachru, E. Silverstein, {\sl Matrix
Descriptions of (1,0) Theories in Six-Dimensions}, Phys.Lett.{\bf
B420} (1998) 55, hep-th/9709118.

\bibitem{lowe}
D. Lowe, {\sl $E_8 \times E_8$ Small Instantons in Matrix Theory},
Nucl.Phys.{\bf B519} (1998) 180, hep-th/9709015.

\bibitem{johnson}
C. V. Johnson, {\sl Anatomy of a Duality}, Nucl.Phys.{\bf B521}
(1998) 71, hep-th/9711082.

\bibitem{ferrara}
S. Ferrara, A. Kehagias, H. Partouche, A. Zaffaroni, {\sl
Membranes and five-branes with Lower Supersymmetry and their AdS
Supergravity Duals}, Phys.Lett.{\bf B431} (1998) 42,
hep-th/9803109.

\bibitem{tatar}
C. Ahn, K. Oh and R. Tatar, {\sl Orbifolds of $ AdS_7 \times S^4 $
and Six Dimensional $ (0,1) $ SCFT}, hep-th/9804093.

\bibitem{Oz-p}
O. Aharony, Y. Oz, Z. Yin, {\sl M Theory on $ AdS_p \times
S^{11-p} $ and Superconformal Field Theories}, Phys. Lett. {\bf
B430} (1998) 87, hep-th/9803051.


\bibitem{Minwalla-p}
S. Minwalla, {\sl Particles on $ AdS_{4/7} $ and Primary Operators
on $ M_{2/5} $ brane Worldvolumes}, J. High Energy Phys. {\bf
9810} (1998) 002, hep-th/9803053.

\bibitem{leigh}
R. G. Leigh and M. Rozali, {\sl The Large $N$ Limit of the $ (2,0)
$ Superconformal Field Theory}, Phys. Lett. {\bf B431} (1998) 311,
hep-th/9803068.

\bibitem{Halyo-p}
E. Halyo, {\sl Supergravity on $ AdS_{4/7} \times S^{7/4} $ and M
Branes}, J. High Energy Phys. {\bf 04} (1998) 011, hep-th/9803077.

\bibitem{shamiteva}
S. Kachru and E. Silverstein, {\sl 4-D Conformal Theories and
Strings on Orbifolds}, Phys.Rev.Let.{\bf 80} (1998) 4855,
hep-th/9802183.

\bibitem{juanetal}
O. Aharony, A. Fayyazuddin and J. Maldacena, {\sl The Large N
Limit of $ {\cal N} = 2,1 $ Field Theories from Threebranes in
F-theory}, J. High Energy Phys. {\bf 07} (1998) 013,
hep-th/9806159.

\bibitem{berkooz}
M. Berkooz, {\sl A supergravity dual of a $(1,0)$ field theory in
six dimensions}, Phys. Lett. {\bf B430} (1997) 237,
hep-th/9802195.

\bibitem{Horava}
P. Ho\v{r}ava and E. Witten, {\sl Heterotic and Type $I$ string
dynamics from eleven dimensions}, Nucl. Phys. {\bf B460} (1996)
506, hep-th/9510209; P. Horava and E. Witten, {\sl
Eleven-Dimensional Supergravity on a manifold with Boundary},
Nucl. Phys. {\bf B475} (1996) 94, hep-th/9603142.

\bibitem{oscillator}
M. G\"unaydin, P. van Nieuwenhuizen, N. P. Warner, {\sl General
Construction of the Unitary Representations of Anti-de sitter
Superalgebras and the Spectrum of the $S^4$ Compactification of
11-Dimensional Supergravity}, Nucl. Phys. {\bf B255} (1985) 63.

\bibitem{deboer}
J. de Boer, {\sl Six-Dimensional Supergravity on $ S^3 \times
AdS_3 $ and 2d Conformal Field Theory}, hep-th/9806104.

\bibitem{minwalla}
S. Minwalla, {\sl Restrictions Imposed by Superconformal
Invariance on Quantum Field Theories}, hep-th/9712074.

\bibitem{massless}
K. Pilch, P. K. Townsend and P. van Nieuwenhuizen, {\sl
Compactification of $d=11$ supergravity on $S^{4}$ (or $11=7+4$,
too)}, Nucl. Phys. {\bf B242} (1984) 377.

\bibitem{KKspectrum}
P. van Nieuwenhuizen, {\sl The complete mass spectrum of $d\!=\!
11$ supergravity compactified on $S_4$ and a general mass formula
for arbitrary cosets $M_4$}, Class. Quant. Grav. {\bf 2} (1985)~1.

\bibitem{FF}
S. Ferrara, C. Fr{\o}nsdal, {\sl Gauge Fields as Composite
Boundary Excitations}, Phys.Lett.{\bf B433} (1998) 19,
hep-th/9802126

\bibitem{distlerzamora}
J. Distler and F. Zamora, {\sl Nonsupersymmetric Conformal Field
Theories from Stable Anti-De Sitter Spaces}, hep-th/9810206.


\end{thebibliography}
\end{document}